\documentclass[pra,twocolumn,showpacs,preprintnumbers,]{revtex4}
\usepackage{amsmath}
\usepackage{graphicx}
\usepackage{natbib}
\usepackage{latexsym}

\begin{document}

\title{Vortex solitons in dipolar Bose-Einstein Condensates}
\author{I. Tikhonenkov$^a$, B. A. Malomed$^b$, and A. Vardi$^a$}
\affiliation{$^a$Department of Chemistry, Ben-Gurion University of the Negev, P.O.B. 653,
Beer-Sheva 84105, Israel}
\affiliation{$^b$Department of Physical Electronics, School of Electrical Engineering,
Faculty of Engineering, Tel Aviv University, Tel Aviv 69978, Israel}

\begin{abstract}
We predict solitary vortices in quasi-planar condensates of dipolar atoms,
polarized parallel to the confinement direction, with the effective sign of
the dipole-dipole interaction inverted by means of a rapidly rotating field.
Energy minima corresponding to vortex solitons with topological charges ${%
\ell }=1$ and $2$ are predicted for moderately strong dipole-dipole
interaction, using an axisymmetric Gaussian ansatz. The stability of the
solitons with $\ell =1$ is confirmed by full 3D simulations, whereas their
counterparts with $\ell =2$ are found to be unstable against splitting into
a set of four fragments (quadrupole).
\end{abstract}

\maketitle

\section{Introduction}

Matter-wave patterns in Bose-Einstein condensates (BECs) are sustained by
the interplay between the external trapping potential and intrinsic
interactions between atoms. In particular, bright solitons and soliton
trains in nearly one-dimensional (1D) traps \cite{Li-Rb} are supported by
the relatively weak attraction between atoms of $^{7}$Li, or the stronger
attraction in the $^{85}$Rb condensate. For repulsive interactions, adding
an axial optical-lattice (OL) potential gives rise to gap solitons, as
demonstrated in $^{87}$Rb \cite{Markus}.

In planar 2D geometry with intrinsic repulsion, delocalized
vortices constitute basic BEC patterns \cite{vortex}. The creation
of 2D matter-wave solitons (as well as 2D spatiotemporal solitons,
alias ``light bullets", in nonlinear optics \cite{review}) is a
challenge, as the contact attraction leads to collapse in this
case. Square-shaped OL can stabilize fundamental solitons and
solitary vortices (solitons with embedded vorticity) in two
dimensions \cite{BBB,Yang}. As concerns vortex solitons, the
lattice breaks the isotropy of the embedding space and the related
angular momentum conservation. Nevertheless, the
intrinsic topological charge of the vortex can be defined
unambiguously as $\ell \equiv \Delta \Phi /2\pi $, where $\Delta
\Phi $ is the circulation of the phase of the respective complex
wave function around the vortex' pivot. The simplest
``crater-shaped" vortices, in the form of a single density peak
with an inner hole, trapped, essentially, in a single
cell of the square lattice, is unstable \cite{Zyss}. Stable
vortices with charge $\ell $ can be constructed, in the simplest form,
as sets of four peaks, with phase shift $\Delta \varphi
=\pi \ell /2$ between adjacent ones. Two stable four-peak vortex 
structures are possible, ``rhombuses" (with a
nearly empty cell in the middle) \cite{BBB} and more compact
``squares" \cite{Yang}. For $\ell =2$, i.e.  $\Delta \varphi =\pi
$, these patterns are actually quadrupoles. Higher-order
stable vortices, up to $\ell =6$, were found too, in the form of
circular chains of $8$ or $12$ peaks \cite{HS}. Also
found were ``supervortices", built as ring-shaped chains of $%
12$ compact (crater-shaped) local vortices with individual vorticity $l_{0}=1
$, onto which global vorticity $\ell =\pm 1$ is imprinted \cite{HS}.  The supervortex 
is stable despite the instability of crater-shaped vortices in isolation. 
Two-dimensional solitons, as well as
vortices and quadrupoles of the rhombic type, can also be stabilized by the
quasi-1D OLs, i.e., periodic potentials depending on a single coordinate
\cite{BBB3}.

For repulsive contact interactions, square-shaped OL
can support 2D fundamental and vortical gap solitons \cite{GS}. In addition,
axisymmetric radial potentials may stabilize solitons, including vortical
ones, in both cases of the attractive and repulsive interactions \cite{BBB2}.

Despite the theoretical progress, 2D matter-wave solitons have not yet been
observed, vortex solitons being a still more challenging subject. Therefore,
the search for viable settings allowing the realization of such structures
remains highly relevant. In parallel to BEC, theoretical and experimental
studies of multidimensional spatiotemporal solitons draw great interest in
nonlinear optics \cite{review}.

New possibilities for producing 2D solitons emerge in dipolar quantum gases,
such as BECs of magnetically polarized $^{52}$Cr atoms \cite{Cr}, dipolar
molecules \cite{hetmol}, or atoms with electric moments induced by strong
electric field \cite{dc} or laser illumination \cite{lightDD}. Of these systems, 
the gas of chromium atoms is the medium available for current experiments. For
axisymmetric geometry, with dipoles polarized perpendicular to the 2D
plane, the natural anisotropic dipole-dipole (DD) interaction gives rise to
in-plane repulsion and axial attraction, which can support vortex
lattices \cite{Pu} and, in principle, 2D gap solitons (in the presence of
the respective OL). On the other hand, the sign of the DD interaction may be
reversed by means of rapid rotation of the dipoles \cite{reversal} or using
a combination of microwave and dc fields \cite{Zoller}, which enables the
creation of isotropic solitons \cite{Pedri05}. However, in the full 3D
geometry, isotropic vortex lines are destabilized by the DD interactions
\cite{VortexLine}. Alternatively, stable \textit{anisotropic} solitons can
be supported by the natural DD interaction, when dipoles are polarized in
the 2D plane \cite{Tikhonenkov08}. Related work in nonlinear optics employed
the nonlocal thermal nonlinearity to predict stable vortex rings, with
topological charges $\ell =1$ and $3$ \cite{Krolik} and 2D elliptically
shaped spatial solitons \cite{Moti1}.

Here, we assume the axisymmetric configuration as in Ref. \cite{Pedri05},
with the aforementioned sign inversion of the DD interaction \cite{reversal}%
, to predict stable 2D vortex solitons. Families of vortex soliton states,
with $\ell =1$ and $2$, are constructed in the framework of the 3D
Gross-Pitaevskii (GP) equation, including the long-range DD interactions,
contact repulsion, and transverse confinement potential. With topological
charge $\ell =1$, the solitary vortex is stable, whereas the vortex soliton
with $\ell =2$ splits via a quadrupole instability. Since the only reported
experimental demonstration of 2D spatial vortex solitons requires the
presence of photonic lattices in photorefractive crystals \cite{Moti}, and
because no observation of vortex solitons was reported in uniform media, the
proposal to create such solitons in dipolar BECs is pertinent to the
experiment, especially in view of recent advances in tuning out the local
nonlinearity via a Feshbach resonance (FR) \cite{recent}.

The paper is organized as follows. In Section I, we apply a
variational approximation to predict localized-vortex states
yielding a minimum of the dipolar condensate energy.
Naturally, these states have a chance to represent stable vortex
solitons. In Section III, results of direct simulations of vortex
solitons, performed in the framework of the full 3D\ GP\ equation,
are summarized. For that purpose, we use both numerically exact
profiles, which are generated from the variational
\textit{ans\"{a}tze} by means of the preliminarily simulated
propagation in imaginary time, and the \textit{ans\"{a}tze}
themselves, the corresponding results being quite similar. The
paper is concluded by Section IV. In particular, in that section
we discuss the physical significance of three-body
losses induced by the FR.

\section{Variational analysis for energy minima}

We assume that a strong magnetic field aligns dipole moments along
confinement axis $z$ \cite{reversal,Zoller,Pedri05}. The respective GP
energy functional, expressed in terms of BEC order-parameter $\psi \left(
x,y,z\right) $, is
\begin{equation}
E\left\{ \psi \right\} =T+V+U+U_{d},
\end{equation}%
where the kinetic, confinement, and contact-interaction energies are,
respectively,
\begin{equation}
T=\frac{1}{2}\int \left\vert \nabla \psi (\mathbf{r})\right\vert ^{2}d%
\mathbf{r}~,~V=\frac{1}{2}\int z^{2}|\psi (\mathbf{r})|^{2}d\mathbf{r}~,
\end{equation}%
\begin{equation}
U=\frac{g}{2}\int |\psi (\mathbf{r})|^{4}d\mathbf{r}~,
\end{equation}%
and the DD mean-field energy is
\begin{equation}
U_{d}=\frac{g_{d}}{2}\int \int \left[ 1-\frac{3\left( z-z^{\prime }\right)
^{2}}{\left\vert \mathbf{r}-\mathbf{r^{\prime }}\right\vert ^{2}}\right]
\times |\psi (\mathbf{r^{\prime }})|^{2}|\psi (\mathbf{r})|^{2}\frac{d%
\mathbf{r}d\mathbf{r^{\prime }}}{\left\vert \mathbf{r}-\mathbf{r^{\prime }}%
\right\vert ^{3}}~.
\end{equation}%
Here and below, the length, time, and energy are scaled as $\mathbf{r}%
\rightarrow \mathbf{r}/l_{z}$, $t\rightarrow \omega _{z}t$, and $%
E\rightarrow E/\left( \hbar \omega _{z}\right) $, where $\omega _{z}$ is the
transverse-trap frequency and $l_{z}\equiv \sqrt{\hbar /m\omega _{z}}$ is
its respective length. The interaction strengths are $g=4\pi Na_{s}/l_{z}$
and $g_{d}=Nd^{2}m\left( \hbar ^{2}l_{z}\right) ^{-1}$, where $a_{s}>0$ is
the $s$-wave scattering length, $d$ and $m$ the atomic dipole moment and
mass, $N$ the number of atoms, and the normalization is taken in the form
of $\int \left\vert \psi \left( \mathbf{r}\right) \right\vert ^{2}d\mathbf{r}%
=1$.

To approximate vortex-soliton states with topological charge $\ell $, we use
the normalized Gaussian ansatz in cylindrical coordinates $z$, $\rho \equiv
\sqrt{x^{2}+y^{2}}$, and $\phi \equiv \tan ^{-1}(y/x)$,
\begin{equation}
\psi _{\ell }(\rho ,z,\phi )=A_{\ell }\rho ^{\ell }\exp \left( -\left(
\alpha \rho ^{2}+\gamma z^{2}\right) /2\right) \exp (i\ell \phi ),
\label{psi}
\end{equation}%
where $A_{\ell }^{2}=(\pi ^{-3/2}/\ell )\alpha ^{\ell +1}\gamma ^{1/2}$ for $%
\ell =1,2$. Evaluating $E\{\psi _{\ell }\}$, we obtain,
\begin{equation}
E\left\{ \psi _{1}\right\} =\alpha +\frac{1}{4}\left( \gamma +\frac{1}{%
\gamma }\right) +\frac{1}{2}\alpha \sqrt{\frac{\gamma }{2\pi }}\left[ \frac{g%
}{4\pi }+\frac{g_{d}}{3}f_{1}(\kappa )\right] ,  \label{Elo}
\end{equation}%
\begin{equation}
E\left\{ \psi _{2}\right\} =\frac{3}{2}\alpha +\frac{1}{4}\left( \gamma +%
\frac{1}{\gamma }\right) +\frac{3}{8}\alpha \sqrt{\frac{\gamma }{2\pi }}%
\left[ \frac{g}{4\pi }+\frac{g_{d}}{3}f_{2}(\kappa )\right] ,  \label{Elt}
\end{equation}%
where functions
\begin{equation}
f_{1}(\kappa )=-1+3\int_{0}^{1}R(\kappa ,x)\left[ 1+Q^{2}(\kappa ,x)\right]
dx,  \notag  \label{f1}
\end{equation}%
\begin{equation}
f_{2}(\kappa )=-1+3\int_{0}^{1}R(\kappa ,x)\left[ 1+\frac{2}{3}Q^{2}(\kappa
,x)+Q^{4}(\kappa ,x)\right] dx,  \notag  \label{f2}
\end{equation}
\begin{equation}
R(\kappa ,x)\equiv \frac{(\kappa x)^{2}}{(\kappa x)^{2}+(1-x^{2})},~Q(\kappa
,x)\equiv \frac{1-x^{2}}{(\kappa x)^{2}+(1-x^{2})},
\end{equation}%
depend solely on the \textit{aspect ratio}, $\kappa \equiv \sqrt{\gamma
/\alpha }$.

In strongly prolate (cigar-shaped) geometry with $\kappa \ll 1$, one has $%
R\rightarrow 0$, $Q\rightarrow 1$, and $f_{1,2}(\kappa )\rightarrow $ $-1$.
By contrast, for an oblate (pancake) shape with $\kappa \gg 1$, we have $%
R\rightarrow 1$, $Q\rightarrow 0$, and $f_{1,2}(\kappa
)\rightarrow 2$. The change of the sign of $f_{1,2}$ corresponding
to the transition from $\kappa \ll 1$ to $\kappa \gg 1$ is due to
the respective change of the relative strength of ``side-by-side"
and ``head-to-tail" DD interactions, which dominate the prolate
and oblate
configurations, respectively. Consequently, for $g_{d}>0$ and fixed $\gamma $%
, the integrated DD energy, $U_{d}$, \textit{decreases} both for $\alpha
\rightarrow 0$ (since $U_{d}>0$ for large $\kappa $) and $\alpha \rightarrow
\infty $ (because $U_{d}<0$ in this limit), leading to either expansion or collapse 
along $\rho$.

Inversion of the sign of $g_{d}$ \cite{reversal,Zoller,Pedri05},
converts the energy maximum at the prolate/oblate transition point,
into a minimum. It is thus required to stabilize 2D isotropic patterns, with the
dipolar moments polarized along the cylindrical axis. This requirement for 
$g_{d}$ inversion, in combination with the necessity to reduce the strong contact 
repulsion using
the FR, is an experimental challenge. The existence of stable \emph{\
anisotropic} fundamental (non-topological) solitons, with the dipoles
polarized in the $\left( x,y\right) $ plane and the ordinary sign of the DD
interaction ($g_{d}>0$) \cite{Tikhonenkov08}, suggests that anisotropic
topological (vortex-like) solitons may be found in the same setting.
However, we leave the analysis of such complex patterns to a separate work,
aiming here to retain the cylindrical symmetry, adopting the assumption of $%
g_{d}<0$.

Assuming the sign inversion of the DD interaction, the minimum in $U_{d}$ as
a function of $\alpha $ at $\gamma \sim 1$ will translate into a minimum of
total energy $E$, provided that $U_{d}$ is large enough to offset the
contact-interaction and gradient terms, $U$ and $T$, both scaling linearly
with $\alpha $. This requirement results in the necessary condition for the
existence of a stable isotropic vortex soliton,
\begin{equation}
g_{d}<6\sqrt{2\pi }\ell +3g/\left( 4\pi \right) <-2g_{d}.  \label{ness}
\end{equation}%
The inequalities on the right- and left-hand sides of Eq. (\ref{ness})
guarantee, severally, $\partial E(\alpha ,\gamma )/\partial \alpha <0$ for $%
\alpha \rightarrow 0$ ($f_{\ell }\rightarrow 2$) and $\partial E(\alpha
,\gamma )/\partial \alpha >0$ for $\alpha \rightarrow \infty $ ($f_{\ell
}\rightarrow -1$). As expected, conditions (\ref{ness}) can only be
satisfied for $g_{d}<0$. The required strength, $\left\vert g_{d}\right\vert
$, increases with $\ell $ due to the growing centrifugal contribution to the
kinetic energy, which must be balanced by the attractive part of the DD
interaction. In the strong-interaction regime, with $3g/4\pi \gg 6\sqrt{2\pi
}\ell $, the kinetic term may be neglected, and Eq. (\ref{ness}) simplifies
to $-g_{d}/g>3/(8\pi )$.

Condition (\ref{ness}) only guarantees a minimum of the GP energy as a
function of $\alpha $ for fixed $\gamma $, which is not yet sufficient for a
true minimum of $E(\alpha ,\gamma )$ in the $\left( \alpha ,\gamma \right) $
plane. In particular, fixing $\kappa $ and varying $\gamma =\kappa
^{2}\alpha $, one can see that $U+U_{d}\propto \gamma ^{3/2}$, $V\propto
1/\gamma $, and $T\propto \gamma $, so that kinetic energy $T$ cannot
balance interaction terms $U+U_{d}$ for large $\gamma $. Since $U+U_{d}<0$
in the large-$\kappa $ sector of the $\left( \alpha ,\gamma \right) $ plane,
this will lead to the 3D collapse, with $\alpha ,\gamma \rightarrow \infty $%
, \emph{unless} the vertical-confinement size $l_{z}$ is smaller than the
effective healing length, determined by the interplay of the kinetic energy
with the combined contact and DD interactions.

\begin{figure}[tbp]
\centering
\includegraphics[width=0.5\textwidth]{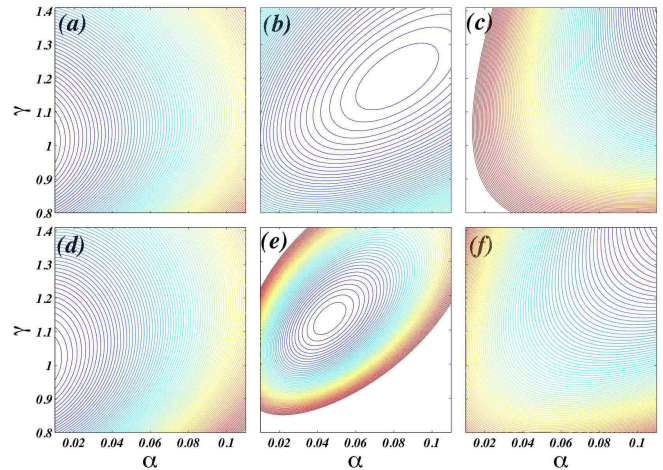}
\caption{(Color online) The GP energy functional for a vortex-soliton ansatz
(\protect\ref{psi}), with $\ell =1$ (a-c) and $\ell =2$ (d-f), as a function
of variational parameters $\protect\alpha $ and $\protect\gamma $. In all
plots, $g=20$, with $g_{d}=-10$ (a), $-20$ (b), $-30$ (c) for $\ell =1$, and
$g_{d}=-20$ (d), $-30$ (e), $-40$ (f), for $\ell =2$.}
\label{fig1}
\end{figure}

In Fig. 1, we plot the GP energies, as given by Eq.~(\ref{Elo}) (panels \ref{fig1}a-c) and
Eq.~(\ref{Elt}) (panels \ref{fig1}d-f), in three characteristic interaction-parameter
regimes. For $\ell =1$ and $2$ respectively, Figs. 1a and 1d display cases
when the DD interaction is too weak to satisfy conditions (\ref{ness}). This
results in a monotonic decrease of the energy as $\alpha \rightarrow 0$ at
fixed $\gamma $, causing the radial expansion of the BEC. On the other hand,
if the DD interaction is too strong, the energy decreases monotonically for
fixed $\kappa $ as $\alpha ,\gamma \rightarrow \infty $, implying the 3D
collapse (Figs. 1c and 1f). In the intermediate regime, the DD interaction
is strong enough to offset the dispersive effect of the contact and kinetic
terms, yet is not excessively strong to induce the 3D collapse. This regime
gives rise to local energy minima, which are found at $g=-g_{d}=20$ for $%
\alpha =0.081,\gamma =1.15$ in Fig. 1(b), and at $g=20,g_{c}=-30$ for $%
\alpha =0.046,\gamma =1.13$ in Fig. 1(e). These minima suggest the
possibility of metastable oblate vortex solitons, with radial widths of $%
3.51l_{z}$ and $4.66l_{z}$ for $\ell =1,2$ respectively.

\section{Mean-field dynamics and stability}

To directly verify the existence and stability of axisymmetric vortex
solitons, we substituted the values of $\alpha $ and $\gamma $ corresponding
to the local energy minima into ansatz (\ref{psi}), and simulated its
evolution according to the full time-dependent GP equation,
\begin{eqnarray}
i\frac{\partial }{\partial t}\psi  &=&\left\{ -\frac{1}{2}\nabla ^{2}+\frac{1%
}{2}z^{2}+g|\psi |^{2}\right.   \label{GPE} \\
~ &~&\left. +g_{d}\int \left[ 1-\frac{3\left( z-z^{\prime }\right) ^{2}}{%
\left\vert \mathbf{r}-\mathbf{r^{\prime }}\right\vert ^{2}}\right] |\psi (%
\mathbf{r^{\prime }})|^{2}\frac{d\mathbf{r^{\prime }}}{\left\vert \mathbf{r}-%
\mathbf{r^{\prime }}\right\vert ^{3}}\right\} \psi ~.  \notag
\end{eqnarray}

\begin{figure}[tbp]
\centering
\includegraphics[width=0.5\textwidth]{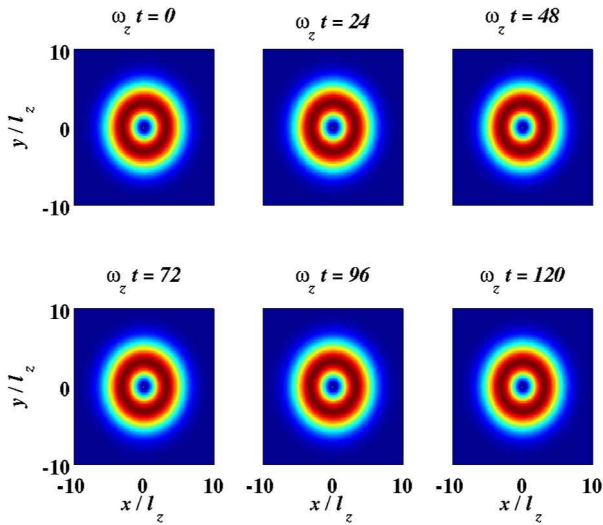}
\caption{(Color online) The evolution of a vortex soliton with $\ell =1$,
for $g=-g_{d}=20$. The initial conditions are obtained by the imaginary-time
propagation for $i\protect\omega _{z}t=30$, starting from Gaussian ansatz (%
\protect\ref{psi}) with $\protect\alpha =0.08$ and $\protect\gamma =1.15$,
which corresponds to the energy minimum in Fig. 1b. The evolution of the
density profile in cross section $z=0$ demonstrates the stability of the
solitary vortex with  $\ell =1$.}
\label{fig2}
\end{figure}

First, the propagation in imaginary time was carried out to reshape the
input into a numerically exact stationary solitary vortex. The amplitude
difference between the initial Gaussian ansatz and reshaped soliton was $%
<10\%$. Then, to test the stability of the solitary vortices, we used these
profiles as initial conditions and carried out 3D simulations in the real
time. The local density and phase at $z=0$, during the real-time evolution
for approximately 20 trap periods, are shown in Figs.~\ref{fig2} and \ref%
{fig3} for $\ell =1$, and in Figs.~\ref{fig4} and \ref{fig5} for $\ell =2$.
With topological charge $\ell =1$, the vortex soliton remains virtually
unchanged during the evolution (Fig.~\ref{fig2}), which demonstrates its
full stability. By contrast, with $\ell =2$, the solitary vortex is
modulationally unstable against splitting into a quadrupole set, as shown in
Fig.~\ref{fig4}. This observation is reminiscent of the stability analysis
for solitary vortices in condensates with the local self-attraction, trapped
in the axisymmetric parabolic potential \cite{VortexStability}, where only $%
\ell =1$ vortices have their stability region, all vortices with $\ell >1$
being inherently unstable.

\begin{figure}[tbp]
\centering
\includegraphics[width=0.5\textwidth]{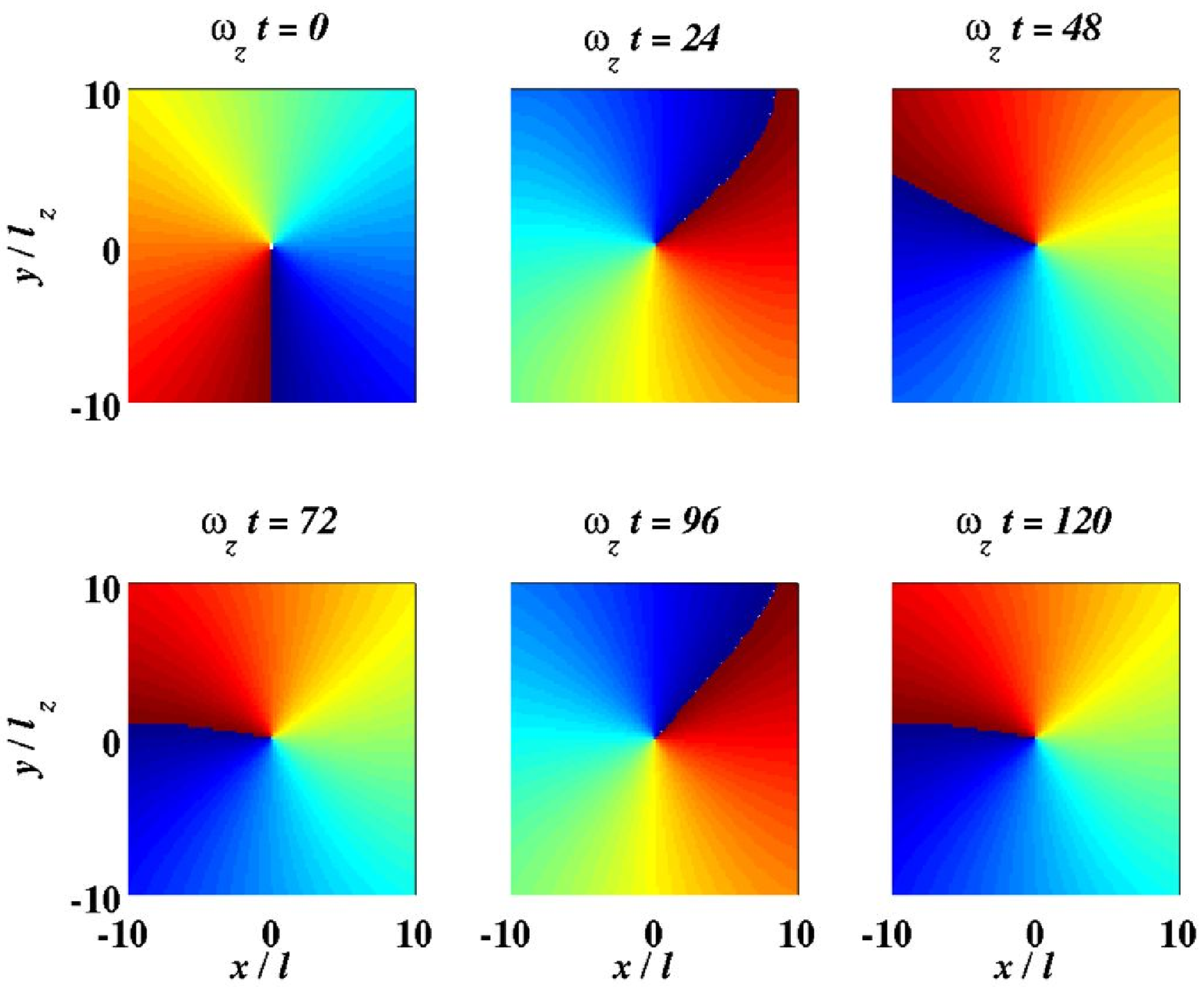}
\caption{(Color online) Phase plots during the evolution of the solitary
vortex with  $\ell =1$ and the same parameters as in Fig. \protect\ref{fig2}
}
\label{fig3}
\end{figure}

\begin{figure}[tbp]
\centering
\includegraphics[width=0.5\textwidth]{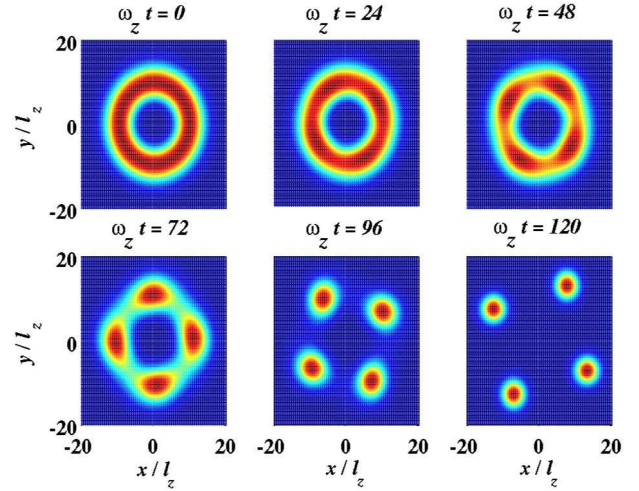}
\caption{(Color online) The evolution of the density profile in cross
section $z=0$ for a vortex soliton with $\ell =2$ and interaction strengths $%
g=20$, $g_{d}=-30$. Initial conditions were generated by the imaginary-time
propagation for $i\protect\omega _{z}t=30$, starting from Gaussian ansatz (%
\protect\ref{psi}) with $\ell =2$, $\protect\alpha =0.046$, $\protect\gamma %
=1.13$, which corresponds to the energy minimum in Fig. 1e. The solitary
vortex is unstable, and eventually splits into a quadrupole set.}
\label{fig4}
\end{figure}

\begin{figure}[tbp]
\centering
\includegraphics[width=0.48\textwidth]{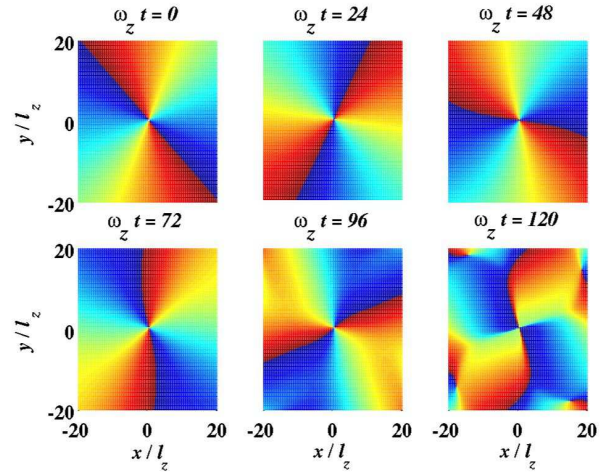}
\caption{(Color online) Phase plots during the evolution of the solitary
vortex with  $\ell =2$ and the same parameters as in Fig. \protect\ref{fig4}
}
\label{fig5}
\end{figure}

\begin{figure}[tbp]
\centering
\includegraphics[width=0.5\textwidth]{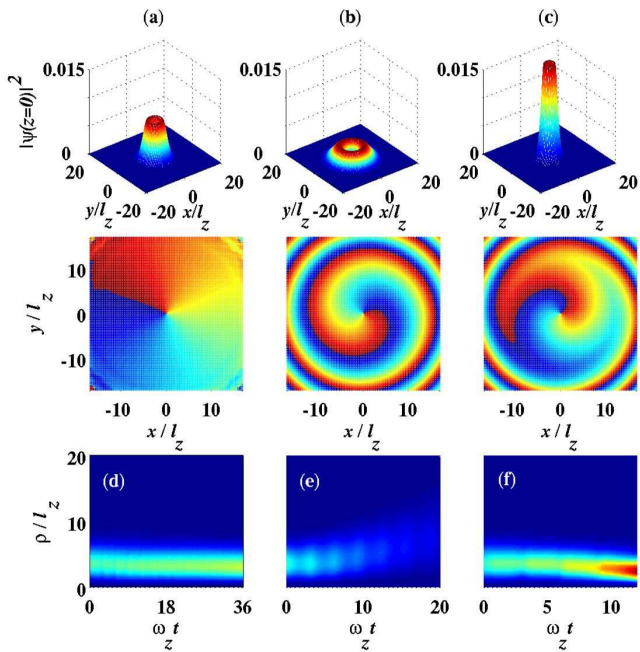}
\caption{(Color online) Density profiles (first line) and phase plots
(second line) in cross section $z=0$, after the real-time propagation for $%
\protect\omega _{z}t=12$, starting from the vortex-soliton state with $\ell
=1$, as taken from Fig. 2. The robust evolution of the solitary vortex for $%
g=-g_{d}=20$ (a) is compared to the expansion for $g=20$, $g_{d}=0$ (b) and
collapse for $g=0$, $g_{d}=-20$ (c). Panels (d)-(f) show the real-time
evolution of the density profile $|\protect\psi (\protect\rho ,z=0,\protect%
\phi )|^{2}$, starting with approximate Gaussian ansatz (\protect\ref{psi}),
for the same parameters.}
\label{fig6}
\end{figure}

The evolution of the solitary vortex with $\ell =1$, was also compared to
the GP dynamics in limit cases when either the contact repulsion or DD
interaction is turned off. As mentioned above, the contact interaction may
be experimentally controlled by means of the FR \cite{recent}, and the DD
interaction may be tuned using additional external fields \cite%
{reversal,Zoller}, or simply turned off by removing the polarizing field,
which allows the dipolar order to frustrate. In Fig.~\ref{fig6}(a-c), we
display the density and phase distributions in cross section $z=0$, for $%
\ell =1$ and $\omega _{z}t=12$. Recall that the vortex soliton was robust
for unaltered interaction strengths (Fig.~\ref{fig6}a). In contrast to that,
when the DD interaction is turned off (Fig.~\ref{fig6}b), radial expansion
is observed, being accompanied by an outgoing density current. On the other
hand, when the $g_{d}/g$ ratio is too large, the vortex collapses (see Fig.~%
\ref{fig6}c), and the current is funneled towards the vertical axis. It is
noteworthy that the axial symmetry is maintained in all cases.

The stability of the solitary vortex with  $\ell =1$ is further corroborated
by the simulated propagation in real time of the initial configuration
corresponding to Gaussian ansatz (\ref{psi}), with $\alpha $ and $\gamma $
set to the location of the minimum in Fig. 1b. This state, which actually
corresponds to a slightly perturbed vortex soliton, demonstrates, for $%
g=-g_{d}=20$, stable self-trapping into the exact solitary vortex (Fig.~\ref%
{fig6}d). By contrast, expansion (Fig.~\ref{fig6}e) and collapse (Fig.~\ref%
{fig6}f) are observed, respectively, in the regimes of too weak (as in Fig.~%
\ref{fig1}a) and too strong (Fig.~\ref{fig1}c) DD interactions.

\section{Conclusions}

The experimental feasibility of quasi-2D solitons in dipolar BECs was
estimated in Refs. \cite{Pedri05} and \cite{Tikhonenkov08}. For $^{52}$Cr
atoms, the natural DD/contact interaction strength ratio is less than $0.1$
\cite{Cr}. Therefore, essential attenuation of the contact interaction by
means of the FR is necessary \cite{recent}. One consequence of this
requirement to the experiment is that three-body losses, which are also
induced by the FR, of order $10^{-28}$ cm$^{6}$ s$^{-1}$ \cite{PfauLosses},
will set an upper limit on the free-evolution time and impose an intrinsic
time dependence. Preliminary numerical simulations including a quintic loss
term of this magnitude indicate that soliton behavior is not considerably
affected by these losses. The remaining difficulty (common with that for
fundamental solitons predicted in Ref. \cite{Pedri05}) is the necessity to
invert the sign of the DD interaction by means of the rapidly rotating
magnetic field \cite{reversal}, a technique which still has to be
experimentally demonstrated.

In conclusion, using the variational analysis and direct simulations of the
GP equation in three dimensions, we have predicted the existence of stable
quasi-2D vortex solitons with topological charge $\ell =1$ in the dipolar
BEC with atomic moments polarized perpendicular to the 2D plane, and the
inverted sign of the dipole-dipole interaction. While energy minima exist
also for solitary vortices with $\ell =2$, the resulting soliton is unstable
and splits into a quadrupole set after the evolution in the course of a few
trap periods. Future work will explore the possibility of anisotropic
solitary vortices with an in-plane polarization axis and the natural sign of
the DD interactions.

\section*{Acknowledgements}

We appreciate valuable discussions with T. Pfau. This work was
supported by the Israel Science Foundation (Grant 582/07).

\end{document}